\documentclass[%
 reprint,
nobibnotes,
 amsmath,amssymb,
 aps,
prl 
]{revtex4-2}

\usepackage{aas_macros}
\usepackage{graphicx}% Include figure files
\graphicspath{{./}{figures/}}
\usepackage{dcolumn}% Align table columns on decimal point
\usepackage{bm}% bold math
\usepackage[usenames,dvipsnames]{xcolor}
\usepackage{hyperref}% add hypertext capabilities
\hypersetup{
	colorlinks=true,
	citecolor=NavyBlue,
	linkcolor=NavyBlue,
	urlcolor=NavyBlue,
}

\usepackage{epsfig}
\usepackage{url}
\usepackage[normalem]{ulem}
\usepackage{latexsym}
\usepackage{epsfig}
\usepackage{amsmath}
\usepackage{amssymb}
\usepackage{wasysym}
\usepackage{graphicx}
\usepackage{verbatim}
\usepackage{enumerate,mdwlist}%lists
\usepackage[titletoc]{appendix}
\usepackage{amsfonts}
\usepackage{pgfplots}
\usepackage[export]{adjustbox}
\usepackage{bbold}

\newcommand{\ba}{\begin{eqnarray}}
\newcommand{\ea}{\end{eqnarray}}
\newcommand{\be}{\begin{equation}}
\newcommand{\ee}{\end{equation}}

\definecolor{grey}{rgb}{0.4,0.4,0.4}
\definecolor{dullmagenta}{rgb}{0.4,0,0.4}
\definecolor{darkblue}{rgb}{0,0,0.4}
\definecolor{midblue}{rgb}{0,0,0.5}
\definecolor{midred}{rgb}{0.5,0,0}
\definecolor{orange}{rgb}{1,0.5,0}
\definecolor{lightbrown}{rgb}{0.75,0.5,0.25}
\definecolor{tan}{cmyk}{0.14,0.42,0.56,0}
\definecolor{djunglegreen}{cmyk}{0.99,0,0.52,0}
\definecolor{lightgreen}{rgb}{0,1,0}
\definecolor{olivegreen}{cmyk}{0.64,0,0.95,0.40}
\definecolor{midgreen}{rgb}{0.0,0.675,0.0}
\definecolor{darkgreen}{rgb}{0,0.5,0}
\definecolor{oxblood}{rgb}{0.5333, 0.0314, 0.0314}

\newcommand{\dingolensing}{\texttt{DINGO-lensing}} 
\newcommand{\dingo}{\texttt{DINGO}} 
\newcommand{\bilby}{\texttt{bilby}}
\newcommand{\modwaveforms}{\texttt{modwaveforms}}
\newcommand{\mur}{$\mu_{\rm rel} \ $}
\newcommand{\delt}{$\Delta t \ $}

\newcommand{\BFlens}{\log_{10}\mathcal{B}_\mathrm{lens}}
\newcommand{\logBFGW}{4.0} 
 
\newcommand{\Xsigma}{4} 
\newcommand{\XsigmaSEOB}{3.4}
\newcommand{\XsigmaXPHM}{3.1} 
 
\newcommand{\Nbackground}{70,000}

\newcommand{\NbackgroundExtra}{140,000} 
\newcommand{\Nforeground}{1,000} 
 
\newcommand{\fractionForegroundLargerGW}{58} 
\newcommand{\fractionForegroundLargerFive}{40} 
\newcommand{\fractionBackgroundLargerZero}{8}

\newcommand{\lettersection}[1]{\textbf{\emph{#1}}.---}

\usepackage[acronym]{glossaries}
\newacronym{GW}{GW}{gravitational wave}
\newacronym{EM}{EM}{electromagnetic} 
\newacronym{GL}{GL}{gravitational lensing}
\newacronym{GO}{GO}{geometric optics} 
\newacronym{WO}{WO}{wave optics}
\newacronym{CBC}{CBC}{compact binary coalescence}
\newacronym{BBH}{BBH}{binary black hole}
\newacronym{BNS}{BNS}{binary neutron star}
\newacronym{NSBH}{NSBH}{neutron-star black-hole binary}
\newacronym{SIE}{SIE}{singular isothermal ellipsoid}
\newacronym{SNR}{SNR}{signal-to-noise ratio}
\newacronym{PE}{PE}{parameter estimation}
\newacronym{LVK}{LVK}{LIGO--Virgo--KAGRA}

\makeatletter
\newcommand*{\glsplainhyperlink}[2]{%
  \colorlet{currenttext}{.}% store current text color
  \colorlet{currentlink}{\@linkcolor}% store current link color
  \hypersetup{linkcolor=currenttext}% set link color
  \hyperlink{#1}{#2}%
  \hypersetup{linkcolor=currentlink}% reset to default
}
\let\@glslink\glsplainhyperlink
\makeatother

\usepackage{orcidlink}

\begin{document}

\title{Discovering gravitational waveform distortions from lensing: \\
A deep dive into GW231123}
\author{Juno C.~L.~Chan %\orcidlink{}
}
\email{chun.lung.chan@nbi.ku.dk}
\author{Jose Mar\'ia Ezquiaga %\orcidlink{}
}
\email{jose.ezquiaga@nbi.ku.dk}
\author{Rico K.~L.~Lo %\orcidlink{}
}
\email{kalok.lo@nbi.ku.dk}
\author{Joey Bowman %\orcidlink{https://orcid.org/0009-0003-3582-1984}
}
\author{Lorena Maga\~na Zertuche %\orcidlink{0000-0003-1888-9904}
}
\author{Luka Vujeva %\orcidlink{}
}
\affiliation{Center of Gravity, Niels Bohr Institute, Blegdamsvej 17, 2100 Copenhagen, Denmark}

\date{\today}

\begin{abstract}
Gravitational waves (GWs) are unique messengers 
as they travel through the Universe without alteration 
except for gravitational lensing. 
Their long wavelengths make them susceptible to diffraction by cosmic structures, providing an unprecedented opportunity to map dark matter substructures. 
Identifying lensed events requires the analysis of thousands to millions of simulated events to reach high statistical significances.
This is computationally prohibitive with standard GW parameter estimation methods. 
We exploit \dingolensing, a deep-learning algorithm that accelerates the inference from CPU days to minutes to thoroughly reanalyze GW231123, the most promising lensing candidate to date.
By performing more than 200,000 simulations with three different waveform models, we find that its statistical significance is below $\Xsigma\sigma$ and the event cannot be claimed as lensed.
We observe that \fractionBackgroundLargerZero\% of GW231123-like nonlensed simulations favor lensing, which could be explained by the self-similarity of short-duration signals. 
Still, \fractionForegroundLargerGW\% of GW231123-like lensed simulations have larger support for lensing,  showing that higher detection statistics are possible. 
We show that analyzing simulations with different waveform models only lowers the significance, highlighting the relevance of waveform systematics. 
Although GW231123 exposes the challenges of claiming the first GW lensing detection, our deep-learning methods have demonstrated to be powerful enough to enable the upcoming discovery of lensed GWs. 
\end{abstract}

\maketitle

\glsresetall
\lettersection{Introduction}%  
\Glspl{GW} travel through the Universe unimpeded with the exception of lensing. 
Unlike optical telescopes, which have a limited field of view and typical cadences larger than  
a day, \gls{GW} observatories are monitoring the entire sky at all times---barring maintenance---with an exquisite time resolution of milliseconds. This gives us the sensitivity to a broad range of lens masses from $\sim1M_\odot$ to $10^{15}M_\odot$. It is predicted that the first strongly lensed \gls{GW} detection could happen in upcoming observing runs~\cite{Ng:2017yiu,Oguri:2018muv,Xu:2021bfn,Wierda:2021upe,Smith:2022vbp}. 
Such a discovery  
opens new frontiers to 
probe gravity in novel ways~\cite{Goyal:2020bkm,Ezquiaga:2020dao,Goyal:2023uvm,Finke:2021znb,Chung:2021rcu}, shed light onto the nature of dark matter~\cite{Tambalo:2022wlm,Fairbairn:2022xln,Vujeva:2025kko,Vujeva:2025nwg}, and unveil new populations of black holes~\cite{Madau:2001sc,Sasaki:2018dmp} and lenses~\cite{Gais:2022xir,Basak:2021ten,Urrutia:2021qak,Barsode:2024wda}. 

The \gls{LVK} detectors are sensitive to \glspl{GW} produced by stellar-mass compact binary coalescences
~\cite{LIGOScientific:2014pky,VIRGO:2014yos,KAGRA:2020tym}. 
Their wavelengths can be comparable to the size of 
isolated black holes, dark matter subhalos or other compact lenses.  
Therefore, the signal could be diffracted and the potentially repeated chirps could interfere with each other as dictated by wave-optics lensing~\cite{Takahashi:2003ix}. 
The lensed signals could be detectable in current searches with lens masses of $\sim 10M_\odot - 10^5M_\odot$~\cite{Chan:2024qmb}, and this imprints distinct changes in the waveform morphology that become smoking-gun signatures of lensing \cite{Lo:2024wqm}. 

However, due to subtle changes in the signal, lensing candidates need to be scrutinized against data quality issues, such as noise fluctuations and seismic disturbance, as well as waveform systematics~\cite{Keitel:2024brp}. To claim a detection at high statistical significance, e.g., $3\sigma$ or $5\sigma$, requires understanding the detection statistic, which, in turn, demands simulating and analyzing thousands to millions of nonlensed events, respectively. 
With present \gls{GW} lensing analysis codes taking on average $\sim50$ CPU days to analyze a single event~\footnote{Private communication with M. Wright, developer of \texttt{Gravelamps} 
\cite{Wright:2021cbn}.}, this would be computationally intractable. 
This is exemplified in the latest \gls{LVK} catalog, GWTC-4.0~\cite{GWTC4},  
which has brought the most massive and highly spinning binary, GW231123\_135430 (GW231123), that can also be interpreted as the highest ranked lensed candidate to date~\cite{GW231123}. 
The lensed hypothesis has been thoroughly studied by \gls{LVK}~\cite{GWTC4_lensing}, finding the event to be an outlier. 
Associated waveform uncertainties, potential noise artifacts, and low prior odds complicate a clear interpretation. 
Independent analyses also find large support for lensing~\cite{Goyal:2025eqo,Hu:2025lhv}, and similar explanations  such as overlapping signals~\cite{Hu:2025lhv}. 
Nonetheless, a robust assessment of the statistical significance of GW231123's lensing interpretation is missing.

The limited number of observable GW cycles in high-mass events allows for alternative interpretations that are difficult to disentangle. 
GW190521~\cite{GW190521,GW190521properties} is another example of an event investigated in a broad range of scenarios from eccentricity~\cite{Romero-Shaw:2020thy} to dynamical captures~\cite{Gamba:2021gap}, head-on collisions of Proca stars~\cite{CalderonBustillo:2020fyi}, and primordial black holes~\cite{DeLuca:2020sae}. 
With a merger rate density of $\sim0.1\,\mathrm{Gpc^{-3}\mathrm{yr}^{-1}}$ for GW190521/GW231123-like signals~\cite{GW231123,GW190521properties}, 
we expect to observe more events of this kind in the next observing runs.
Developing robust strategies to infer source parameters and to quantify the statistical significance for competing interpretations is, therefore, essential.

\gls{GW} parameter estimation has recently experienced a revolution, thanks to deep-learning techniques. 
In particular, neural posterior estimation~\cite{Green:2020dnx,Green:2020hst,Dax:2021tsq} is able to drastically reduce the inference time from order of days down to minutes by avoiding the expensive likelihood evaluations needed by conventional methods~\cite{Ashton:2018jfp}.
This capability enables the assessment of statistical significance for subtle astrophysical effects, such as orbital eccentricity~\cite{Gupte:2024jfe}, which was previously not feasible due to the prohibitive computational cost of traditional sampling algorithms. 
In this \emph{Letter}, we demonstrate that deep-learning techniques are essential for the discovery of lensed \glspl{GW}. 
Building on top of the state-of-the-art algorithm, \dingo{}~\cite{Green:2020dnx,Dax:2021tsq,Dax:2021myb,Wildberger:2022agw,Dax:2022pxd}, we developed \dingolensing~\cite{Chan:2025pdf}, which we exploit to thoroughly reanalyze GW231123 and assess its statistical significance for the first time~\footnote{Other implementations of lensed parameter estimation with deep-learning have been proposed in parallel~\cite{Caldarola:2025oxr,Su:2025xry}.}. 
Our results and methods lay out the detection strategies of lensed \glspl{GW} for upcoming observing runs.

\lettersection{Gravitational waveform distortions from lensing}% 
Starting from an emitted \gls{GW} signal $h(t)$, the observed lensed waveform $h_{\rm L}(t)$ is obtained by propagating $h(t)$ through the curved spacetime defined by the lens. 
Assuming linear perturbations, weak-field gravity, and a lens much smaller than the source--observer distance, this propagation is conveniently solved in Fourier space. 
Lensing acts like a transfer function between the frequency-domain emitted $\tilde h(f)$ and observed signals $\tilde h_{\rm L}(f)$, $F(f)=\tilde h_{\rm L}(f)/\tilde h(f)$, where the amplification function $F(f)$ is defined by the diffraction integral~\cite{Takahashi:2003ix},
\begin{equation} \label{eq:diffraction_integral}
    F(f|\vec{\theta}_s)=\frac{\tau_D f}{i}\int \mathrm{d}^2\theta\, e^{2\pi i f t_d(\vec\theta,\vec\theta_\mathrm{S})}\,,
\end{equation}
which integrates the time delay $t_d$ accumulated by the lensed signal over all paths from source to observer and accounts for their possible interference and diffraction. 
Here $\vec\theta_\mathrm{S}$ is the source position, $\vec\theta$ is the angular position on the lens plane, and 
$\tau_D=\frac{D_{\rm L} D_{\rm S}}{c D_{\rm LS}}$ is a reference time related to the angular diameter distances to the lens $D_{\rm L}$, the source $D_{\rm S}$, and between the lens and the observer $D_{\rm LS}$, respectively.
The frequency dependence of $F(f)$ produces the characteristic waveform distortions in $h_{\rm L}(t)$.

The nature of these distortions depends on the lens profile and the source-lens configuration. 
Recent studies have developed efficient numerical methods to solve the diffraction integral for different lens classes~\cite{Wright:2021cbn,Cheung:2024ugg,Villarrubia-Rojo:2024xcj}. 
Observationally, the strongest distortions arise when repeated chirps from a strongly lensed \gls{GW} source interfere~\cite{Lo:2024wqm,Ezquiaga:2025gkd}, while in the deep wave-optics regime diffraction effects are mild, with $\lim_{f\to0}F(f)=1$~\cite{Takahashi:2003ix}. 
This interference regime can be modeled accurately using the stationary phase approximation~\cite{Ezquiaga:2025gkd}, yielding
\begin{equation}
\label{eq:F_two_images_point}
F(f>0) =  1 + \sqrt{\mu_{\rm rel}}\, e^{i(2\pi f \Delta t - \pi/2)}\,,
\end{equation}
where \delt and \mur are the time delay and relative magnification between two repeated signals, respectively. 
Focusing on the most likely lensing configurations, we assume the second chirp has a lower magnification and opposite parity~\cite{Blandford:1986zz}, i.e. \delt$>0$, \mur$<1$, and a $\pi/2$ phase difference \footnote{Our parametrization differs from the phenomenological model in Ref.~\cite{Liu:2023ikc}, where the number of repeated chirps and their phase differences are treated as free parameters.}. 
The overall magnification is omitted, as it is degenerate with luminosity distance. 
This parametrization encompasses point and fold singularities, as well as many near-cusp configurations~\cite{Lo:2024wqm,Ezquiaga:2025gkd,Serra:2025kbw}. 
Folds and cusps are the most common caustics in generic lenses~\cite{Blandford:1986zz}, while point singularities appear in axisymmetric lenses such as point-mass lenses~\cite{Schneider:1992bmb} or Navarro-Frenk-White profiles~\cite{Navarro:1996gj} that can be associated with compact lenses and dark matter subhalos, respectively. 
Moreover, this simple parametrization has been shown to efficiently recover simulated signals lensed by point-mass lenses~\cite{Chan:2025pdf}.  
Negative-parity single images, which may cause small additional distortions~\cite{Dai:2017huk,Ezquiaga:2020gdt}, are excluded here and left for future work.

Under these assumptions, a \gls{GW} is identified as \textit{lensed} when the inference yields \mur $>0$ and \delt$>0$. 
More complex (and rare) cases beyond Eq.~\eqref{eq:F_two_images_point}, such as overlapping images of the same type with a phase different from $\pi/2$  
or \mur $> 1$ due to substructure~\cite{Vujeva:2025kko,Vujeva:2025nwg}, will be explored in follow-up work.

\lettersection{Accelerating lensed parameter estimation}%
To accelerate \gls{GW} parameter estimation, we employ simulation-based inference, in which a neural network is trained on simulated lensed signals embedded in detector noise to approximate the posterior distribution over the underlying physical parameters.
Details of \dingolensing{} can be found in the companion paper~\cite{Chan:2025pdf} with code at~\setcounter{footnote}{10}\footnote{\href{https://github.com/dingo-lensing/dingo-lensing}{https://github.com/dingo-lensing/dingo-lensing}}. 
The network architecture follows Ref.~\cite{Dax:2021tsq} with identical hyperparameters.
It combines an embedding network that compresses the input strain into $128$ latent features and a normalizing flow that generates the posterior.
Overall, the model contains on the order of $10^8$ trainable parameters. 
The noise is modeled as Gaussian during the training.

We train the lensed networks with $\Delta t \subset (0,0.1)$ sec, and $\mu_{\text{rel}} \subset (0,1)$
\footnote{These priors are chosen as broad phenomenological distributions covering the parameter space relevant for waveform distortions in the LVK sensitive frequency band, rather than as astrophysical population priors. Adopting astrophysical priors can modify the Bayes factor distribution, but their impact can be assessed a posteriori through reweighting without retraining the neural networks.}.
The nonlensed parameters follow standard choices for the extrinsic parameters and spins (with magnitudes up to 0.99) \cite{Chan:2025pdf}. 
In total there are 17 parameters, which is the largest dimensionality achieved with \dingo{} so far. 
Lensed waveforms in \dingolensing{} and \bilby{} are implemented through \modwaveforms~\footnote{\href{https://github.com/ezquiaga/modwaveforms}{https://github.com/ezquiaga/modwaveforms}}. 
We validate the networks as described in \cite{Chan:2025pdf}, testing the recovery of the injected values, comparing the inferences for representative examples with \bilby~\cite{Ashton:2018jfp}, and checking the sampling efficiencies when performing importance sampling~\cite{Dax:2022pxd}. 
\dingolensing\, drastically reduces the computing time compared to conventional methods, sampling with neural posterior estimations in seconds and computing evidences with importance sampling in minutes.
In comparison, point-mass lens analyses of real events with \texttt{Gravelamps}~\cite{Wright:2021cbn}, a code used by \gls{LVK}, take on average $\sim50$ CPU days. 

%-----------------
%FIGURE
%-----------------
\begin{figure}[b!]
    \centering
    \includegraphics[width=0.5\textwidth]{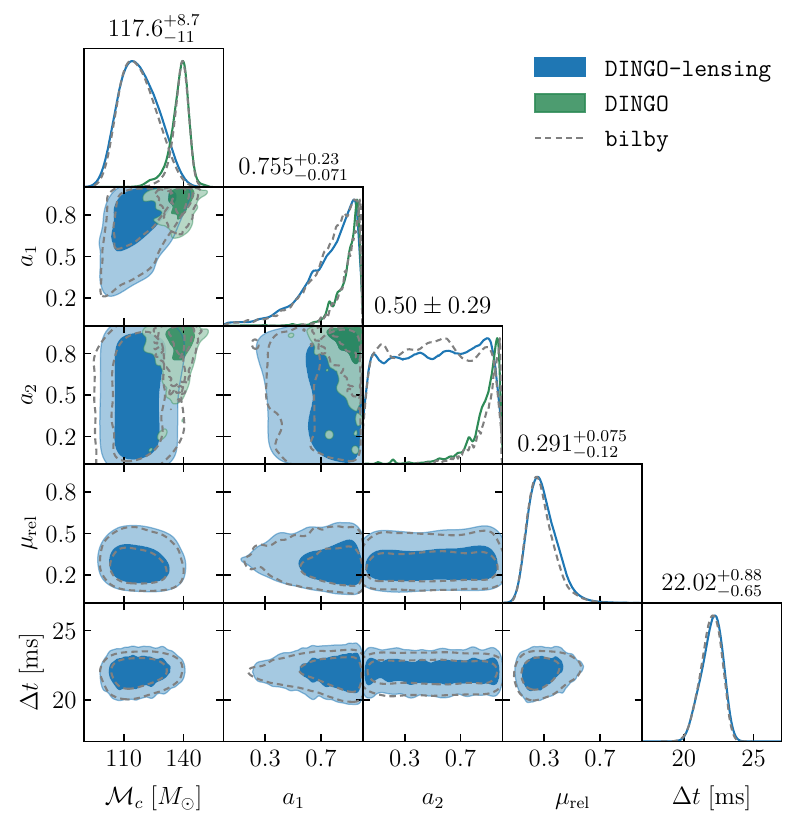}
    \caption{Source characterization of GW231123 assuming it is lensed (with \dingolensing{}, in blue) and not lensed (with \texttt{DINGO}, in green), respectively. 
    From the 17 parameters in the inference, we present the chirp mass $\mathcal{M}_c$, spin magnitudes $a_{1,2}$, lensing relative magnification $\mu_{\text{rel}}$ and time delay $\Delta t$. 
    The posterior distributions using \dingolensing{} and \texttt{DINGO} are compatible with those obtained using \bilby{}.
    The credible levels quoted here are from the \dingolensing{} analysis.
    }
    \label{fig:GW231123_posteriors}
\end{figure}
%-----------------
%-----------------

\lettersection{GW231123: reanalyzing a puzzling lensed candidate}% 
Searches for lensing distortions have been conducted since the first \gls{GW} observing run~\cite{Hannuksela:2019kle,LIGOScientific:2021izm,LIGOScientific:2023bwz,Janquart:2023mvf,Janquart:2024ztv,GWTC4_lensing}. 
However, it has not been until the discovery of GW231123, the most massive binary observed so far, that a strong support was found~\cite{GW231123}. 
This support for lensing was found for a point-mass lens model~\cite{GWTC4_lensing} as well as a point mass embedded in external shear~\cite{Goyal:2025eqo} and holds across different waveform approximants. 
Although Ref.~\cite{GWTC4_lensing} compared the evidence of GW231123 to 254 simulations of nonlensed events following the GWTC-4.0 astrophysical population models~\cite{GWTC4_pop} and found it to be an outlier, the specific properties of this event demand a detailed investigation of GW231123-like signals.    

%-----------------
%FIGURE
%-----------------
\begin{figure}[t!]
    \centering
    \includegraphics[width=\columnwidth]{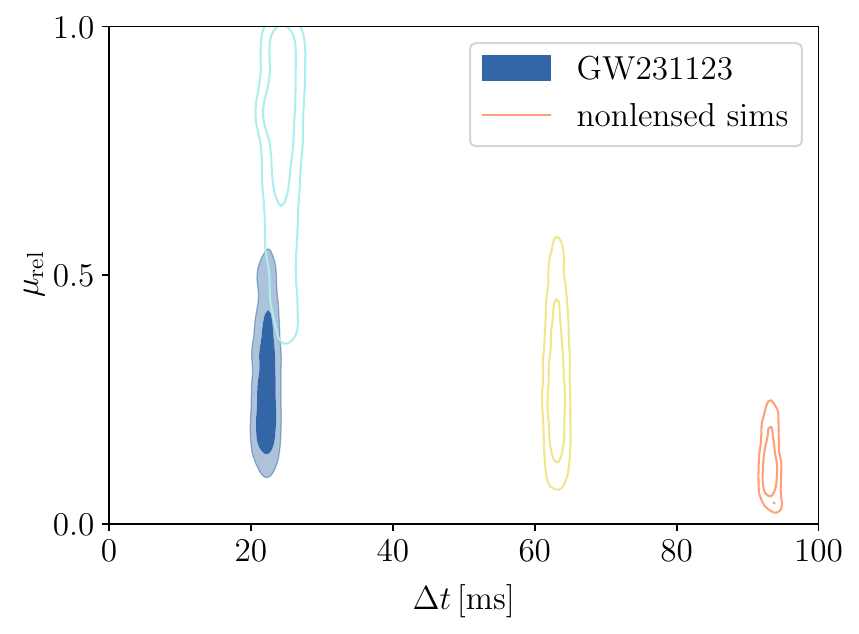}
    \caption{Inference for the lensing relative magnification $\mu_{\text{rel}}$ and time delay $\Delta t$. 
    We compare GW231123 results (filled contours) with nonlensed simulations for similarly heavy binaries (unfilled contours).}
    \label{fig:GW231123_and_simulations}
\end{figure}
%-----------------
%-----------------

For the first time, we train neural networks to learn (non)lensed signals using a numerical relativity surrogate waveform model, \texttt{NRSur7dq4}~\cite{Varma:2019csw}, with a frequency range of $[20, 512]~\rm{Hz}$, and an analysis duration of $8~\rm{sec}$~\footnote{We set the waveform starting frequency at 0 Hz. This means that the model will generate waveforms as far back in time from the merger as allowed.}. \texttt{NRSur7dq4} is the waveform model with the lowest mismatch compared to some GW231123-like numerical relativity simulations~\cite{GW231123}.
We train the networks for chirp masses in the range of $[30,180]M_\odot$, mass ratios in $[1/6,1]$, and luminosity distances in $[0.5,10]~\rm{Gpc}$.
The prior distribution of the remaining source parameters follow Table 1 in~\cite{Chan:2025pdf}, with extended constraints in the component masses. 
We use the average noise power spectral density of the LIGO detectors around the time of the event~\cite{GW231123zenodo}, 
and stop the training at epoch 300 when the loss function plateaus. 

Analyzing the publicly available LIGO Livingston and Hanford data around a reference GPS time $t_{\rm ref} = 1384782888.6$s~\footnote{\href{https://gwosc.org/eventapi/html/O4_Discovery_Papers/GW231123_135430/v1/}{Gravitational Wave Open Science Center}}, our results are presented in Fig.~\ref{fig:GW231123_posteriors} for a subset of the 17D parameter space including the chirp mass, spin magnitudes, and the two lensing parameters. 
For reference, we compare the \dingolensing{} posterior distributions with those obtained with \bilby~\cite{Ashton:2018jfp}. 
The \bilby{}~analysis took $\sim 14$ CPU days, compared to $\sim 32$ CPU \emph{minutes} with \dingolensing{}~(all single-core execution times and scale inversely with the number of cores used).
Apart from the visual agreement, we compare the log Bayes factors $\BFlens$ from the two codes, with a difference of only 0.14. 
When using importance sampling, the sampling efficiency is about $0.58\%$.
At face value, the Bayes factor comparing the lensed and nonlensed hypotheses gives strong support for lensing, $\BFlens=\logBFGW$, and is larger than the ones obtained with the point-mass lens model with a surrogate waveform~\cite{GWTC4_lensing,Goyal:2025eqo}. This difference is expected because the two analyses employ different parametrizations and prior distributions for the lensing model. Nevertheless, both approaches consistently find strong support for lensing relative to the nonlensed hypothesis.

To evaluate the significance of the preference for lensing, we simulate GW231123-like events with component masses uniformly drawn between chirp masses of 90$M_\odot$ and 160$M_\odot$, mass ratios of 0.2 and 1, and luminosity distances uniformly in comoving volume between 0.6 and 8 Gpc within the detector network's range~\cite{KAGRA:2013rdx}. 
Spins are sampled over all possible directions and magnitudes up to 0.99. 
We require the Hanford-Livingston network signal-to-noise ratio to be larger than 8. 
Notably, we find that many instances 
of nonlensed events display posterior distributions in the lensing parameters that deviate from 0 as shown in Fig.~\ref{fig:GW231123_and_simulations} for a selection of them. 
The distance of the lensing posteriors away from 0 can be used as a preliminary indicator of lensing, obtained only with neural posterior estimation in a few seconds~\cite{Chan:2025pdf}.  

%-----------------
\begin{figure}[t!]
\begin{center}
\includegraphics[width=\columnwidth]{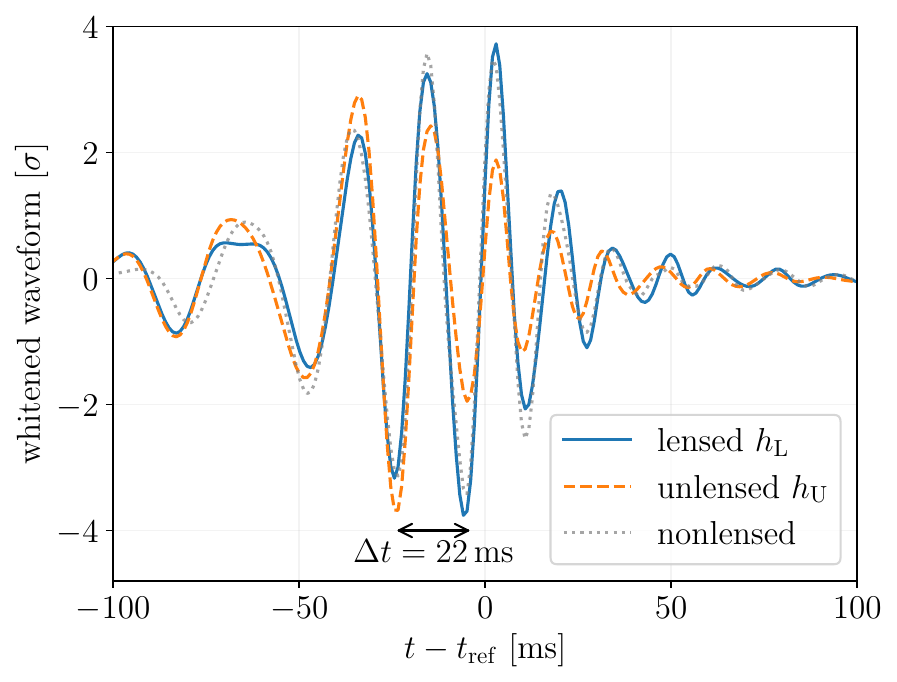}
\end{center}
\caption{\label{fig:GW231123_selfsimilar}Whitened best-fit lensed ($h_{\rm L}$, solid), unlensed ($h_{\rm U}$, dash), and nonlensed ($h$, dotted) time-domain waveforms ($\sigma$) for GW231123. The inferred time delay between the two chirps $\Delta t$ matches with their instantaneous period near the reference time $t_{\rm ref}$.}
\end{figure}
%-----------------

We observe that, in many other instances, the recovered $\Delta t$ is multimodal, and the peaks correlate with the period of the waveform.  
This is due to the self-similarity of short waveforms such as GW231123, for which only a few cycles are observable.
We illustrate this using GW231123 in Fig.~\ref{fig:GW231123_selfsimilar}, where we show the inferred whitened best-fit lensed, unlensed (i.e., \emph{un}doing the lensing amplification from the lensed waveform), and nonlensed (i.e., assuming that there is no lensing) waveforms, respectively \footnote{The whitening uses the estimated noise power spectral density for GW231123 from the LIGO Hanford detector \cite{GW231123}.}.
Our inferred time delay between the two chirps $\Delta t \approx 22$ ms matches with the instantaneous period of the waveforms near the merger at around the reference time $t_{\rm ref}$.

\begin{figure}[t!]
    \centering
    \includegraphics[width=\columnwidth]{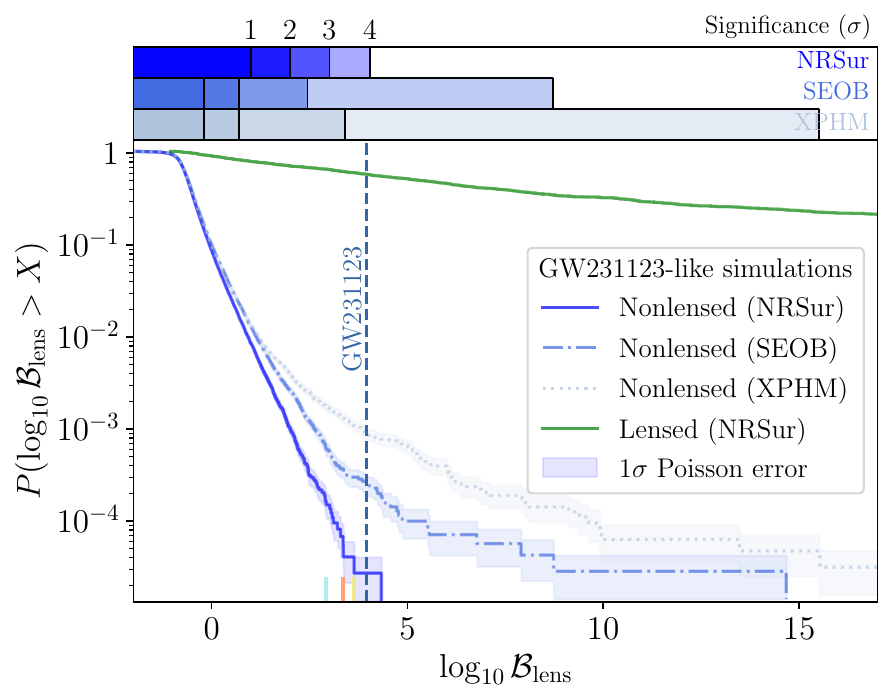}
    \caption{Complementary cumulative distribution function $P(x>X)$ of the logarithmic lensing Bayes factors ($x=\log_{10}\mathcal{B}_\mathrm{lens}$) for GW231123-like simulations. 
    The $\BFlens$ of GW231123, \logBFGW, is indicated by the vertical dashed line. 
    The shaded region corresponds to the Poisson error of $N$ simulations, $1/\sqrt{N}$.  
    Small vertical lines indicate the Bayes factors of the examples displayed in Fig. \ref{fig:GW231123_and_simulations} and the bars on top highlight the false-alarm probability levels of the nonlensed simulations in units of standard deviation ($\sigma$). 
    With our neural networks trained on \texttt{NRSur7dq4}, we analyze nonlensed simulations generated with \texttt{NRSur7dq4} (NRSur), \texttt{SEOBNRv5PHM} (SEOB), and \texttt{IMRPhenomXPHM} (XPHM) waveform approximants.
    }
    \label{fig:backgrounds}
\end{figure}

Furthermore, we compare the GW231123 Bayes factor with a set of more than \Nbackground\, GW231123-like nonlensed events. 
Its complementary cumulative distribution function is plotted in Fig.~\ref{fig:backgrounds}~\footnote{We have validated representative examples of Fig.~\ref{fig:backgrounds} computing $\BFlens$ with standard sampling methods (\bilby), finding a good agreement (within $\lesssim0.5$).}. 
\fractionBackgroundLargerZero\% of nonlensed events favor lensing,  $\BFlens>0$, with a tail that extends to large Bayes factors. 
We find that GW231123 has a $\Xsigma\sigma$ false-alarm probability. 
This is to be contrasted with a set of \Nforeground\,  GW231123-like lensed simulations. 
There, \fractionForegroundLargerGW\% of events have larger Bayes factor than GW231123. 
Despite the clear difficulties to achieve high detection statistics in events like GW231123, 
we project that $\sim$\fractionForegroundLargerFive\% of our lensed simulations would favor lensing with a significance larger than $5\sigma$.

We explore the effects of waveform systematics by analyzing more than \NbackgroundExtra{} additional GW231123-like nonlensed simulations generated with effective-one-body models (\texttt{SEOBNRv5PHM}~\cite{Ramos-Buades:2023ehm}) and inspiral-merger-ringdown phenomenological approximants (\texttt{IMRPhenomXPHM}~\cite{Colleoni:2024knd}), see Fig.~\ref{fig:backgrounds}. 
We find that their inferred distributions of Bayes factors are skewed to larger values, lowering the overall significance of GW231123 to \XsigmaSEOB{}$\sigma$ and \XsigmaXPHM{}$\sigma$, respectively. 

\lettersection{Towards the discovery of lensed gravitational waves}% 
The study of GW231123 serves as a testing ground to prepare the detection strategies of lensed \glspl{GW}. 
We envision a two step process. 
As a first step, we propose an identification of lensed candidates with networks trained on average detector configurations and source properties. 
The goal is to have a flexible machine-learning model that could identify rapidly and efficiently lensing candidates for follow-up. 
Candidate's identification can be done with \dingolensing{} either by checking if the lensing posteriors are away from zero, cf. Fig. \ref{fig:GW231123_and_simulations}, or by computing the Bayes factor, which then needs to be compared against simulations, cf. Fig. \ref{fig:backgrounds}. 

%-----------------
%FIGURE
%-----------------
%-----------------
%----------------- 

As a second step, we propose to train a focused network on the most significant candidates, as we have done for GW231123. 
This is a crucial step in order to capture the nonstationarity of detector noise and the particularities of any given event, for example, waveform systematics, the composition of the detector network, and any data quality issue. 
Multiple networks could be trained to cover different choices.  
The focused networks serve to compute dedicated backgrounds to assess the statistical significance of a particular lensed candidate.
\dingolensing{} training and inference times 
are orders of magnitude more efficient computationally than simulating a background with traditional parameter estimation methods. 

Importantly, the significance of lensing waveform distortions can increase in different scenarios. 
In particular, for strong lensing, if repeated copies of the same signal are identified, and some of them display additional waveform modifications, those will be less likely to be mimicked by other effects. 
This scenario is expected when there are populations of microlenses~\cite{Diego:2019lcd,Mishra:2023ddt} or compact dark matter subhalos near critical curves~\cite{Vujeva:2025nwg}. 

\lettersection{Conclusion}%
The discovery of lensed \glspl{GW} will be groundbreaking, and as such, it requires extraordinary evidence. 
Because of the long wavelengths of \glspl{GW}, the effect of lensing can be imprinted in characteristic waveform distortions. 
Those distortions with respect to the fiducial waveform models can be singled out using Bayesian model comparison. 
However, current lensed parameter estimation methods are computationally too costly to assess the false-alarm probability of interesting candidates to the required statistical significance. 

Taking advantage of recent advances in \gls{GW} parameter estimation using deep learning with \dingo{}~\cite{Green:2020dnx,Green:2020hst,Dax:2021tsq}, we exploit \dingolensing{}~\cite{Chan:2025pdf} to thoroughly reanalyze GW231123, the highest ranked lensed candidate to date.
We find that the event cannot be claimed as lensed: the false-alarm probability of the apparent support for lensing is bounded below $\Xsigma\sigma$, and could be explained by the self-similarity of the signal. 
We show that simulating signals with different waveform approximants only lowers the significance, highlighting the relevance of waveform systematics.
%Other potential sources of false evidence of lensing include waveform systematics and non-stationary noise features~\cite{GWTC4_lensing}. 
%That said, these would only increase the false alarm probability for the support. Our method thus gives us the upper bound on the statistical significance in this scenario. 

We demonstrate that \dingolensing{} is able to efficiently assess the statistical significance of lensed candidates by analyzing large sets of (non)lensed events, a necessary step for the discovery of lensed \glspl{GW}. 
Our work also opens the door to rapid identification of lensed candidates for electromagnetic follow-up. 
This is particularly relevant for the science of multimessenger lensing with binary neutron stars~\cite{Smith:2025axx}.

Although we have focused our analyses on searching for lensed signals, the methods and strategies presented here are equally applicable to the identification of waveform modifications from other scenarios, e.g. modified gravity environmental effects or overlapping signals. 
Our results demonstrate that deep-learning accelerated analysis will play a key role in the discovery of new phenomena in \gls{GW} astronomy.

\hfill\begin{acknowledgments}
\emph{Acknowledgments}.---We are grateful to Gregorio Carullo, Srashti Goyal, Stephen Green, Nihar Gupte, David Keitel and Miguel Zumalac\'arregui for useful comments and discussions along this project. 
We thank Justin Janquart for the internal \gls{LVK} review. 
We also thank Mick Wright for providing the computing cost of Gravelamp. 
This work was supported by Research Grants No.~VIL37766 and No.~VIL53101 from Villum Fonden and the DNRF Chair Program Grant No.~DNRF162 by the Danish National Research Foundation.
This work has received funding from the European Union's Horizon 2020 research and innovation programme under the Marie Sklodowska-Curie Grant Agreement No.~101131233.  
J.~M.~E is also supported by the Marie Sklodowska-Curie Grant Agreement No.~847523 INTERACTIONS.
The Center of Gravity is a Center of Excellence funded by the Danish National Research Foundation under grant no.~DNRF184.
The Tycho supercomputer hosted at the SCIENCE HPC center at the University of Copenhagen was used for supporting this work. 
The authors also acknowledge the extensive use of the Jolene workstation for this work.

This research has made use of data or software obtained from the Gravitational Wave Open Science Center (gwosc.org), a service of the LIGO Scientific Collaboration, the Virgo Collaboration, and KAGRA. This material is based upon work supported by NSF's LIGO Laboratory which is a major facility fully funded by the National Science Foundation, as well as the Science and Technology Facilities Council (STFC) of the United Kingdom, the Max-Planck-Society (MPS), and the State of Niedersachsen/Germany for support of the construction of Advanced LIGO and construction and operation of the GEO600 detector. Additional support for Advanced LIGO was provided by the Australian Research Council. Virgo is funded, through the European Gravitational Observatory (EGO), by the French Centre National de Recherche Scientifique (CNRS), the Italian Istituto Nazionale di Fisica Nucleare (INFN) and the Dutch Nikhef, with contributions by institutions from Belgium, Germany, Greece, Hungary, Ireland, Japan, Monaco, Poland, Portugal, Spain. KAGRA is supported by Ministry of Education, Culture, Sports, Science and Technology (MEXT), Japan Society for the Promotion of Science (JSPS) in Japan; National Research Foundation (NRF) and Ministry of Science and ICT (MSIT) in Korea; Academia Sinica (AS) and National Science and Technology Council (NSTC) in Taiwan.
\end{acknowledgments}

\emph{Data availability}.---\dingolensing{} is available on GitHub~\cite{Note11}. 
The neural networks used in this work will be made public upon publication. 

\bibliography{references}
\end{document}